\documentclass[pra,amsmath,amssymb,showpacs,superscriptaddress, twocolumn]{revtex4}

\usepackage{graphicx}
\usepackage{dcolumn}
\usepackage[hypertex]{hyperref}
\usepackage{bm}
\usepackage[usenames,dvipsnames]{color}
\usepackage{amsmath}
\usepackage{amsfonts}
\usepackage{amsthm}
\usepackage{amssymb}
\usepackage{mathrsfs}
\usepackage{subfigure}
\usepackage{ulem}

\begin{document}

\title{Detecting macroscopic quantum coherence with a cavity optomechanical system}

\author{Qiang Zheng}
\affiliation{School of Mathematics, Guizhou Normal University, Guiyang, 550001, China}
\affiliation{Beijing Computational Science Research Center, Beijing, 100084, China}

\author{Jianwei Xu}
\affiliation{College of Science, Northwest A\&F University, Yangling, Shaanxi 712100, China}

\author{Yao Yao}
\email{yaoyao@mtrc.ac.cn}
\affiliation{Microsystems and Terahertz Research Center, China Academy of Engineering Physics, Chengdu, Sichuan 610200, China}

\author{Yong Li}
\email{yongli@csrc.ac.cn}
\affiliation{Beijing Computational Science Research Center, Beijing, 100084, China}
\affiliation{Synergetic Innovation Center of Quantum Information and Quantum Physics, University of Science and Technology of China,
Hefei, Anhui 230026, China}
\affiliation{Synergetic Innovation Center for Quantum Effects and Applications,   Hunan Normal University, Changsha 410081, China}

\date{\today}

\begin{abstract}
The rigorous resource framework of quantum coherence has been set up recently and excited a wide variety of interests. Here we show that a quantum cavity optomechanical system, as an emerging platform, can behave with a certain value of quantum coherence at a macroscopic scale. We also find that the difference between the total optomechanical coherence and the sum of the optical and the mechanical coherence just equals their mutual information. Motivated by the detection of the optomechanical entanglement, an experimentally feasible scheme to probe the optomechanical coherence is proposed.
\end{abstract}


\maketitle

\section{INTRODUCTION}
Coherence is a fundamental feature in quantum theory
\cite{Dirac58}. It is a necessary condition for quantum correlation, for instance quantum discord \cite{JJMA2016, yao2015}, and quantum entanglement \cite{Streltsov2015}. From the application point of view, coherence is a useful resource for quantum information process \cite{Nielsen00}, quantum metrology \cite{Maccone2014}, thermodynamics
\cite{Correa2014, SWLISUN0215}, biological process \cite{Collini2010}, and asymmetry of quantum states \cite{Spekkens2014}. Historically, a breakthrough theory of quantum coherence is developed in quantum optics \cite{Glauber63, Sudarshan63}. Recently, a coherence theory has been established by regarding it as a physical resource \cite{Plenio2014, Mintert2014, Marvian2013}. Spirited by the rapid development of quantum coherence theory \cite{linori2012, Parthasarathy2016, luo2012, guolilam2015, Winter2016, Adesso2016b, vedral2016, Rastegin2016}, some important questions naturally arise: (1) what's the behavior of the coherence in a realistic system at a macroscopic level and how to experimentally detect the macroscopic coherence? (2) whether there is a relationship between the coherence and quantum mutual information? The aim of this paper is to address these questions.

In principle, no fundamental law prevents the application of
quantum mechanics in a macroscopic system. Many physical systems display the distinct quantum coherence at a macroscopic scale, such as Bose-Einstein condensate \cite{Anderson1995}, Josephson junction \cite{younori2005}, a hybrid system composed by a superconducting qubit and a mechanical oscillator \cite{Cleland2010}, etc.

Cavity optomechanics \cite{Aspelmeyer2012a} as an emerging field, provides us an essentially new platform to probe the quantum coherence in a macroscopic system. In a typical cavity optomechanical system, the motion of the mechanical resonator couples to a single- mode Fabry-P$\acute{e}$rot cavity through radiation pressure. Here the mechanical resonator can be considered as a macroscopic system as it usually involves billions of atoms~\cite{Aspelmeyer2014a}. The important progresses in this realm include (but not limited to) cooling the mechanical resonator to the ground state \cite{Rae2007}, preparing an entanglement state \cite{LTYDWANG2013, Barzanjeh2012a} or a squeezed state \cite{Kronwald2013, Wollman2015, Agarwal2015}, probing the mechanical zero-point fluctuation \cite{Painter2012}, and observing the radiation pressure shot-noise~\cite{purdy2013}.

In this paper, we investigate the macroscopic quantum coherence of
the cavity optomechanical system. We find that: (1) with the increase of the single-photon optomechanical coupling strength, both the mechanical coherence and the total optomechanical coherence are strengthened, while the
optical coherence is suppressed; (2) three kinds of the coherence mentioned above are very sensitive to the cavity decay, but are robust to the decay of the mechanical resonator; (3) the difference between the total optomechanical coherence and the sum of the optical and the mechanical coherence just equals their quantum mutual information; (4) the macroscopically optomechanical coherence can be detected in an all-optical setup.

The structure of this paper is as follows. We explore a standard cavity optomechanical system by the linearization approximation in Sec. II. In Sec. III we investigate the effect of the system parameters, such as the
optomechanical coupling strength and the cavity decay rate, on the
coherence of the cavity optomechanics. We also suggest an experimentally
available all-optical setup to detect the optomechanical coherence.
Finally, in the last section a conclusion is given.
Main aspects of the resource theory for the Gaussian state are reviewed in Appendix A.

\section{Model of the cavity optomechanical system}
We consider a standard cavity optomechanical system, with the Hamiltonian written as ($\hbar=1$ in the following) \cite{jqliao2016, shbar15, Zhang2014a}
\begin{equation}
\hat{H}_{sys}=\Delta_{0} \hat{a}^{\dag} \hat{a}+ \frac{\omega_{m}}{2}(\hat{p}_{0}^2+\hat{q}_{0}^2)-G_{0}\hat{a}^{\dag} \hat{a} \hat{q}_{0}+i E(\hat{a}^{\dag}-\hat{a})
\label{hamA}
\end{equation}
in the rotating frame with respect to the frequency of the driving laser, $\omega_{d}$. Here $\hat{a}$ ($\hat{a}^{\dag}$) is the annihilator (creation) operator of the cavity field. $\hat{q}_{0}\equiv\hat{q}/\sqrt{2}$ and $\hat{p}_{0}\equiv\hat{p}/\sqrt{2}$ are the dimensionless position and momentum of the mechanical mode with frequency $\omega_m$ satisfying the commutation $[\hat{q}_{0}, \hat{p}_{0}]=i$. $\Delta_{0}=\omega_{c}-\omega_{d}$ is the detuning between the cavity mode and the external driving laser, $G_{0}$ is the single-photon optomechanical coupling strength, and $E$ denotes the amplitude of the external driving laser.

The Heisenberg-Langevin equation of the optomechanical system is given as \cite{walls2008}
\begin{subequations}
\label{moeA}
\begin{eqnarray}
\frac{d \hat{q}}{dt} &=& \omega_{m} \hat{p},
\\
\frac{d \hat{p}}{dt} &=& -\omega_{m} \hat{q}-\gamma_{m} \hat{p}+ \sqrt{2}G_{0}\hat{a}^{\dag} \hat{a} + \sqrt{2}\hat{\xi},
\\
\frac{d \hat{a}}{dt} &=& -(\kappa+i \Delta_{0})\hat{a}+ \frac{iG_{0}}{\sqrt{2}}\hat{a}\hat{q}+E+\sqrt{2\kappa} \hat{a}_{\textrm{in}},
\end{eqnarray}
\end{subequations}%
where $\kappa$ is the cavity decay rate, and $\hat{a}_{\textrm{in}}$ denotes the vacuum input noise with the only nonzero correlation \cite{pzoller00, Clerk2011}
\begin{equation}
\langle \hat{a}_{\textrm{in}}(t) \hat{a}_{\textrm{in}}^{\dag}(t') \rangle=\delta(t-t'),
\end{equation}
and $\hat{\xi}$ is the Brownian stochastic force with the zero-mean value and the correlation
$\langle \hat{\xi}(t) \hat{\xi}(t') \rangle=\frac{\gamma_{m}}{\omega_{m}} \int \frac{d \omega}{2 \pi} e^{-i\omega (t-t')}\omega [\coth(\frac{\hbar \omega}{2 k_{B}T})+1]$. In the limit of the high mechanical quantity factor $Q_{m}=\omega_{m}/\gamma_{m} \gg 1$ with $\gamma_m$ being the mechanical damping rate, the Brownian noise $\hat{\xi}$ becomes an approximate Markovian noise
\cite{Benguria1981}
\begin{equation}
\langle \hat{\xi}(t) \hat{\xi}(t') + \hat{\xi}(t') \hat{\xi}(t) \rangle/2 \simeq \gamma_{m}(2 n_{th}+1) \delta(t-t'),
\end{equation}
where the mean thermal phonon number $n_{\textrm{th}}= \{\exp[\hbar \omega_{m}/(k_{B}T)]-1\}^{-1}$ with $T$ the environmental temperature and $k_{B}$ the Boltzmann constant.

With the strong driving laser for the cavity field, as a good approximation, the whole optomechanical system can be described by a stable steady state and the linearized fluctuations around the steady state. The solutions of the steady state are easily obtained by setting Eq.~(\ref{moeA}) to zeros as \cite{xuli2015, DengXTX2016, yli2012a}
\begin{subequations}
\begin{eqnarray}
p_{s} &=& 0,
\\
q_{s} &=& \sqrt{2} G_{0} |\alpha_{s}|^2/\omega_{m},
\\
\alpha_{s} &=& \frac{E}{\kappa+i(\Delta_{0}-G_{0}q_{s}/\sqrt{2} )}.
\end{eqnarray}
\label{moeAB8}
\end{subequations}

Introducing the optical quadratures $\hat{X}=\hat{a}^{\dag}+\hat{a}$ and
$\hat{Y}=i(\hat{a}^{\dag}-\hat{a})$ and the corresponding noises $\hat{X}_{\textrm{in}}$ and
$\hat{Y}_{\textrm{in}}$, the equations for the linearized fluctuations around the steady state can be casted into a compact matrix form
\begin{equation}
\frac{d}{dt} u(t)=\mathbf{A} u(t)+n(t),
\label{moeC}
\end{equation}
with $u(t)=(\delta\hat{q},~\delta\hat{p},~\delta X,~\delta Y)^{\mathrm{T}}$, $n(t)=(0,~\sqrt{2}\hat{\xi},~\sqrt{2\kappa}\hat{X}_{in},~\sqrt{2\kappa}\hat{Y}_{in})^{\mathrm{T}}$, and
\begin{equation}
\mathbf{A}=\left(
\begin{array}{cccc}
0 & \omega_{m}& 0 & 0 \\
-\omega_{m} & -\gamma_{m}& G & 0 \\
0 & 0& -\kappa & \Delta \\
G & 0& -\Delta & -\kappa \\
\end{array}%
\right).
\end{equation}
Here $G=\sqrt{2} G_{0} \alpha_{s}$ and $\Delta=\Delta_{0}-G_{0}q_{s}/\sqrt{2}$. In deriving Eq.~(\ref{moeC}), we assume that $\alpha_{s}$ is a real number, which can be achieved by adjusting the phase of the driving laser.

The formal solution to Eq.~(\ref{moeC}) is~\cite{Meystre2008}
\begin{equation}
u(t)= \mathbf{M}(t)u(0)+\int_{0}^{t} ds \mathbf{M}(s)n(t-s)
\end{equation}
with $\mathbf{M}(t)=\exp(\mathbf{A} t)$. The stability conditions of the matrix $\mathbf{A}$ are derived as $\omega_{m}(\Delta^2+\kappa^2)-G^2\Delta>0$ and $2\gamma_{m}\kappa \{ s_{+}s_{-}+\gamma_{m}[v(\Delta^2+\kappa^2)+2\kappa \omega_{m}^2]\}+\Delta \omega_{m}G^2 v>0$ with $s_{\pm}=\kappa^2+(\omega_{m}\pm \Delta)^2$ and $v=\gamma_{m}+2\kappa$ by making use of the Routh-Hurwitz criteria~\cite{DeJesus1987}.

With the initial Gaussian state of the optomechanical system and the
(approximately) linear dynamics, the state of the system stays as the
Gaussian all the time. For a Gaussian state, it can be fully
characterized by the first moment (which has been obtained in
Eq.~(\ref{moeAB8})) and the second moments or the covariance matrix $\mathbf{V}(t)$
with the element $V_{ij}(t)= \langle u_{i}(t)u_{j}(t)+u_{j}(t)u_{i}(t) \rangle /2$.
In the long-time limit, the formal solution of $V_{ij}(t)$ is written as
\cite{Plenio2008, Paternostro2007}
\begin{equation}
V_{ij}(t)=\int_{0}^{\infty} ds \int_{0}^{\infty}ds' M_{ik}(s) M_{jl}(s') D_{kl}(s-s'),
\label{cmveqA1}
\end{equation}
where the element of the diffusion matrix $\mathbf{D}(t)$ is defined as $D_{kl}(s-s')=\langle n_{k}(s)n_{l}(s')+n_{l}(s)n_{k}(s') \rangle /2$. From Eq.~(\ref{cmveqA1}), the covariance matrix $\mathbf{V}(\infty)$ is determined by the Lyapunov equation
\begin{equation}
\mathbf{A} \mathbf{V}+ \mathbf{V} \mathbf{A}^{T} = -\mathbf{\tilde{D}}
\label{cmveq}
\end{equation}
where $\mathbf{D}(s-s')= \mathbf{\tilde{D}} \delta(s-s')$ with $\mathbf{\tilde{D}}= \mathrm{diag}[0,~2\gamma_{m}(2n_{th}+1),~2\kappa,~2\kappa]$.

\section{Quantum coherence of the optomechanical system}

The analytical solution to Eq.~(\ref{cmveq}) for the total
covariance matrix $\mathbf{V}$ is very cumbersome, so we mainly
adopt the numerical simulations to explore the quantum coherence of
the cavity optomechanical system. This solution can be expressed as
\begin{equation}
\mathbf{V}=\left(
\begin{array}{cc}
\mathbf{V}_{\textrm{mec}} & \mathbf{V}_{\textrm{cor}} \\
\mathbf{V}_{\textrm{cor}}^{\mathrm{T}} & \mathbf{V}_{\textrm{opt}} \\
\end{array}%
\right).
\label{optomecCVB}
\end{equation}
Here $\mathbf{V}_{\textrm{mec}}$, $\mathbf{V}_{\textrm{opt}}$, and
$\mathbf{V}_{\textrm{cor}}$ describe the covariance matrices of the
mechanical resonator, the optical field, and their correlation,
respectively. For the covariance matrix Eq.~(\ref{optomecCVB}), its
two symplectic eigenvalues are
$ \nu_{1,2}= \frac{1}{\sqrt{2}}[ \Gamma \pm \sqrt{\Gamma^2- 4
\mathrm{Det}(\mathbf{V}_{\textrm{tot}}) } ]^{1/2}$
with $\Gamma= \mathrm{Det}(\mathbf{V}_{\textrm{mec}} )+
\mathrm{Det}(\mathbf{V}_{\textrm{opt}} )+2
\mathrm{Det}(\mathbf{V}_{\textrm{cor}} ) $.

According to the definition in Eq.~(\ref{incohstatA2}) or Eq.~(\ref{incohstatA4}), the corresponding quantum coherence can be obtained as%
\begin{equation}
C_{\textrm{mec}} = C(\mathbf{V}_{\textrm{mec}} ),
~ C_{\textrm{opt}}= C(\mathbf{V}_{\textrm{opt}} ),
~C_{\textrm{tot}} = C(\mathbf{V} ).
\label{optomeCoh}
\end{equation}

\begin{figure}[!tb]
\centering
\includegraphics[width=2.5in]{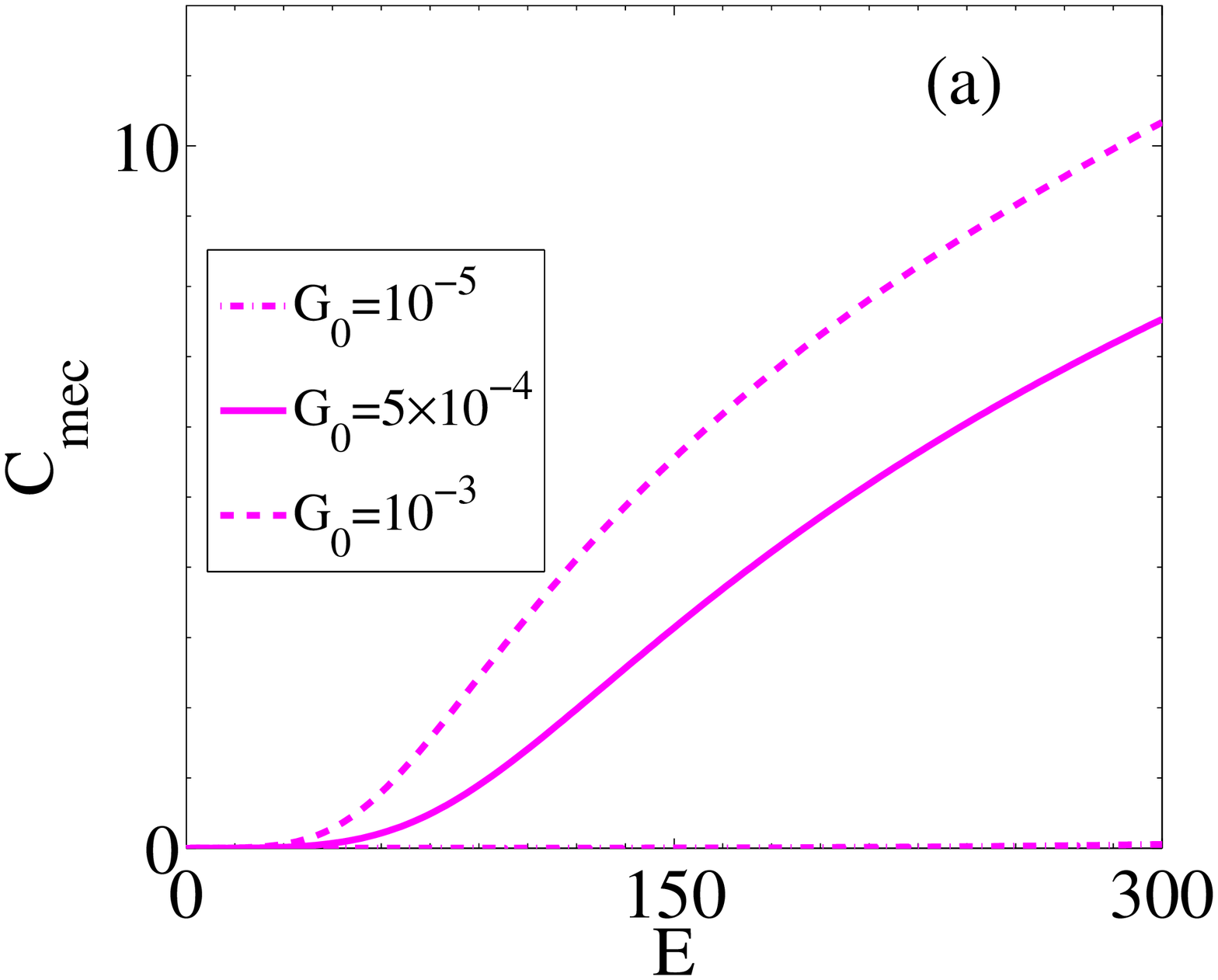}
\includegraphics[width=2.5in]{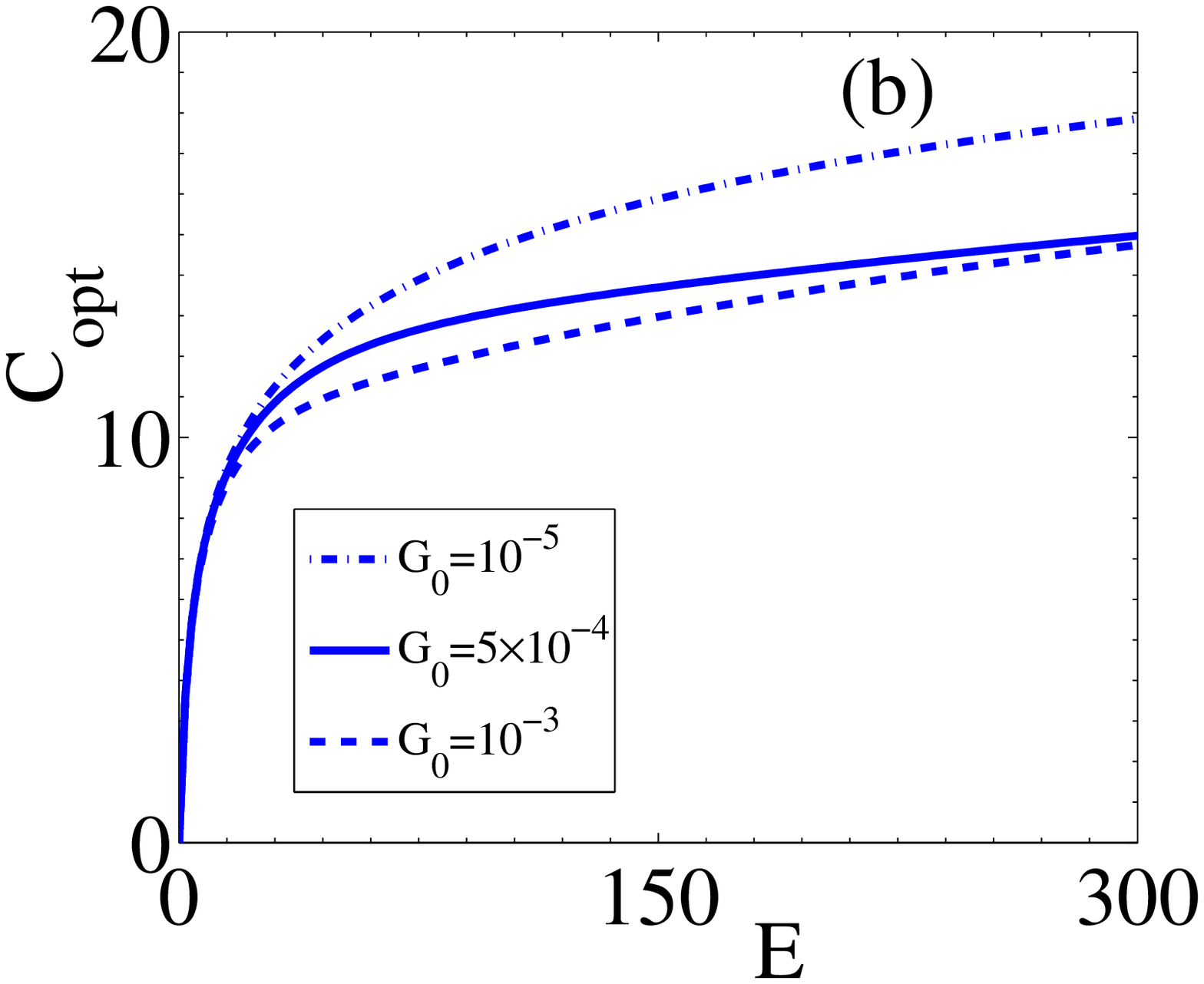}
\includegraphics[width=2.5in]{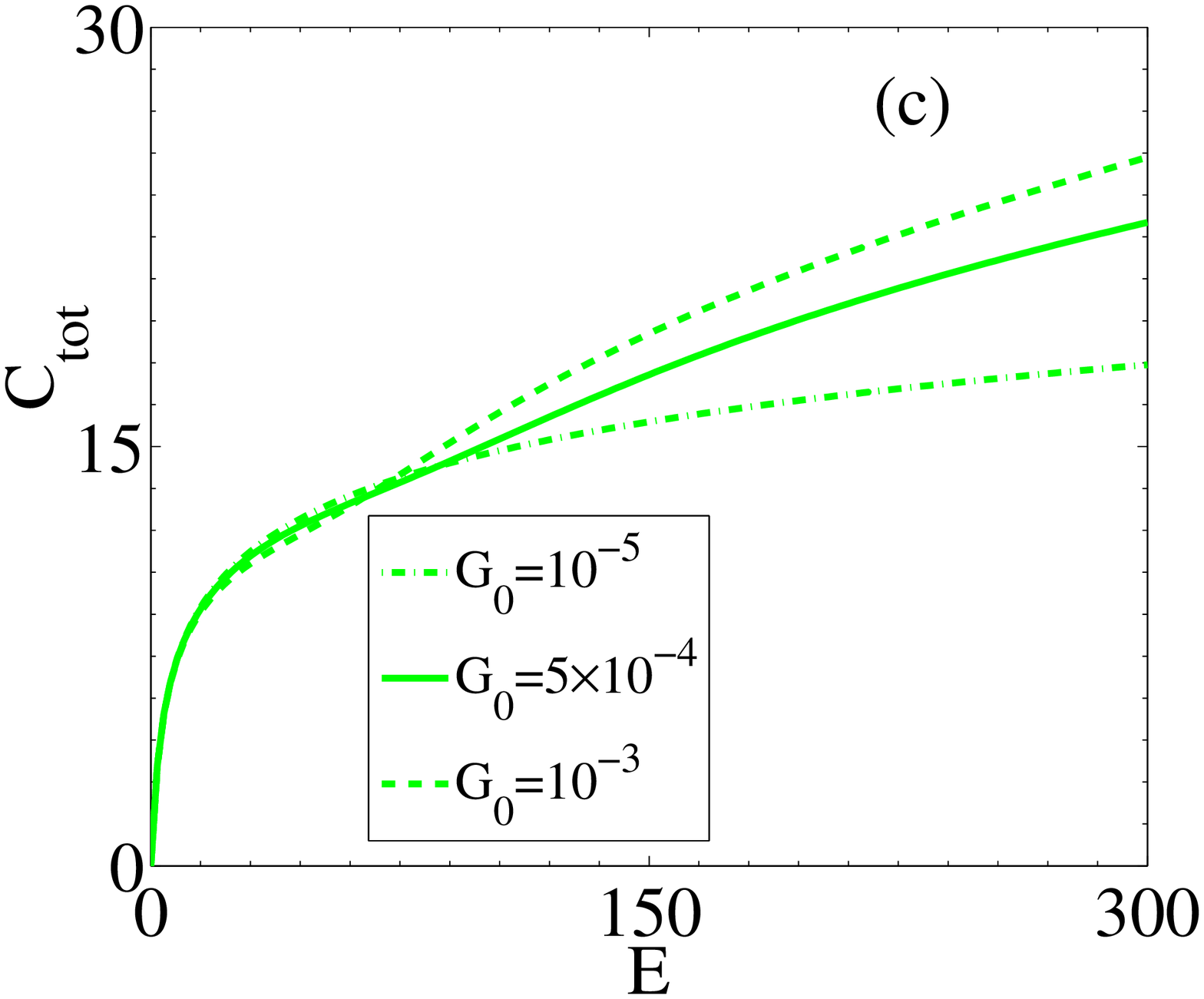}
\caption{(Color online) The quantum coherence as a function of the
driving strength $E$ (scaling with $\omega_{m}=1$). (a), (b), and (c)
correspond to the mechanical coherence $C_{\textrm{mec}}$, the optical coherence $C_{\textrm{opt}}$, and the optomechanical coherence $C_{\textrm{tot}}$, respectively. Other parameters are $\gamma_{m}=0.01\,\omega_{m}$, $\kappa=0.1\,\omega_{m}$, $\Delta_{0}= \omega_{m}$, and $n_{th}=10$.} \label{figB}
\end{figure}

\begin{figure}[!tb]
\centering
\includegraphics[width=2.5in]{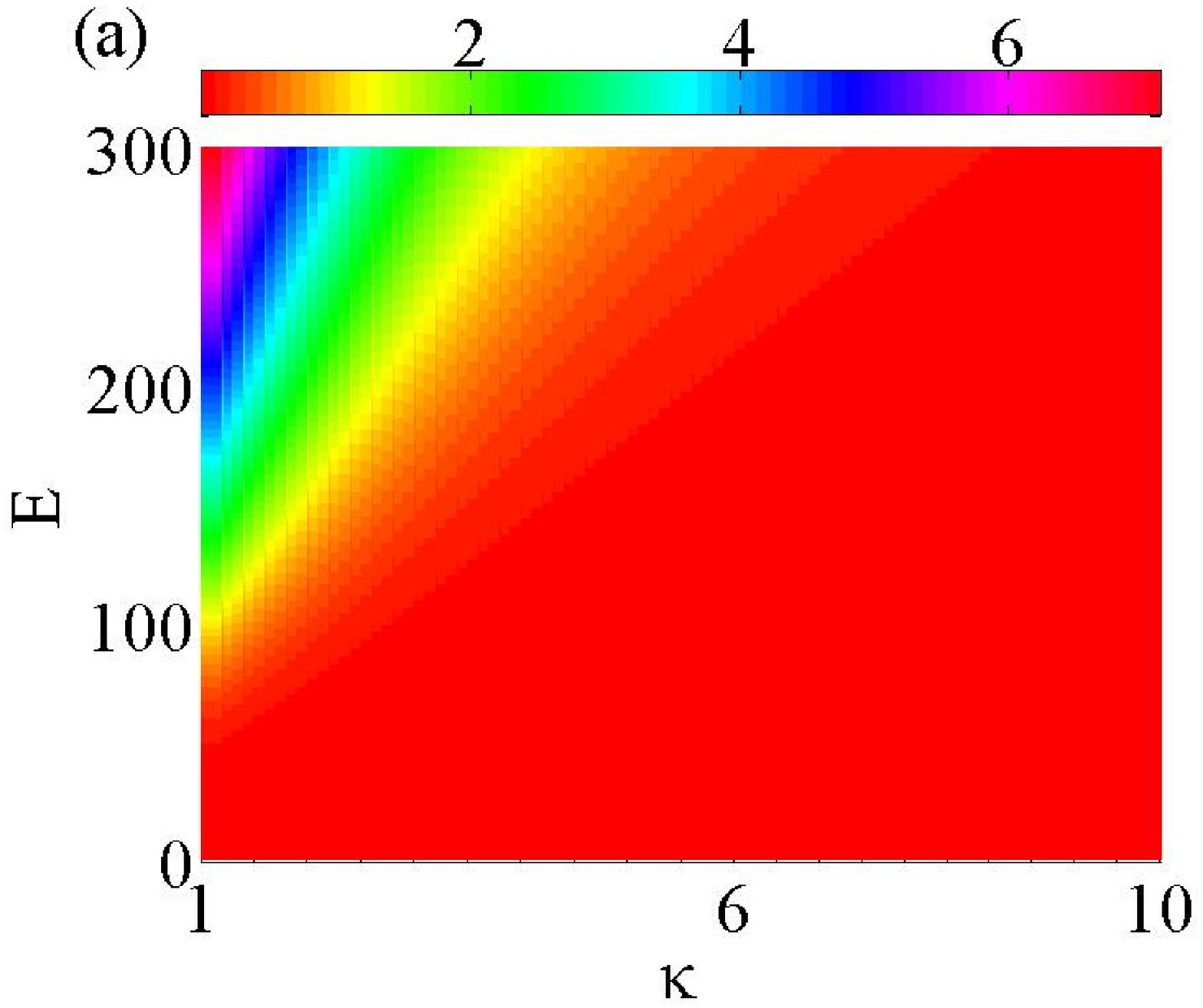}
\includegraphics[width=2.5in]{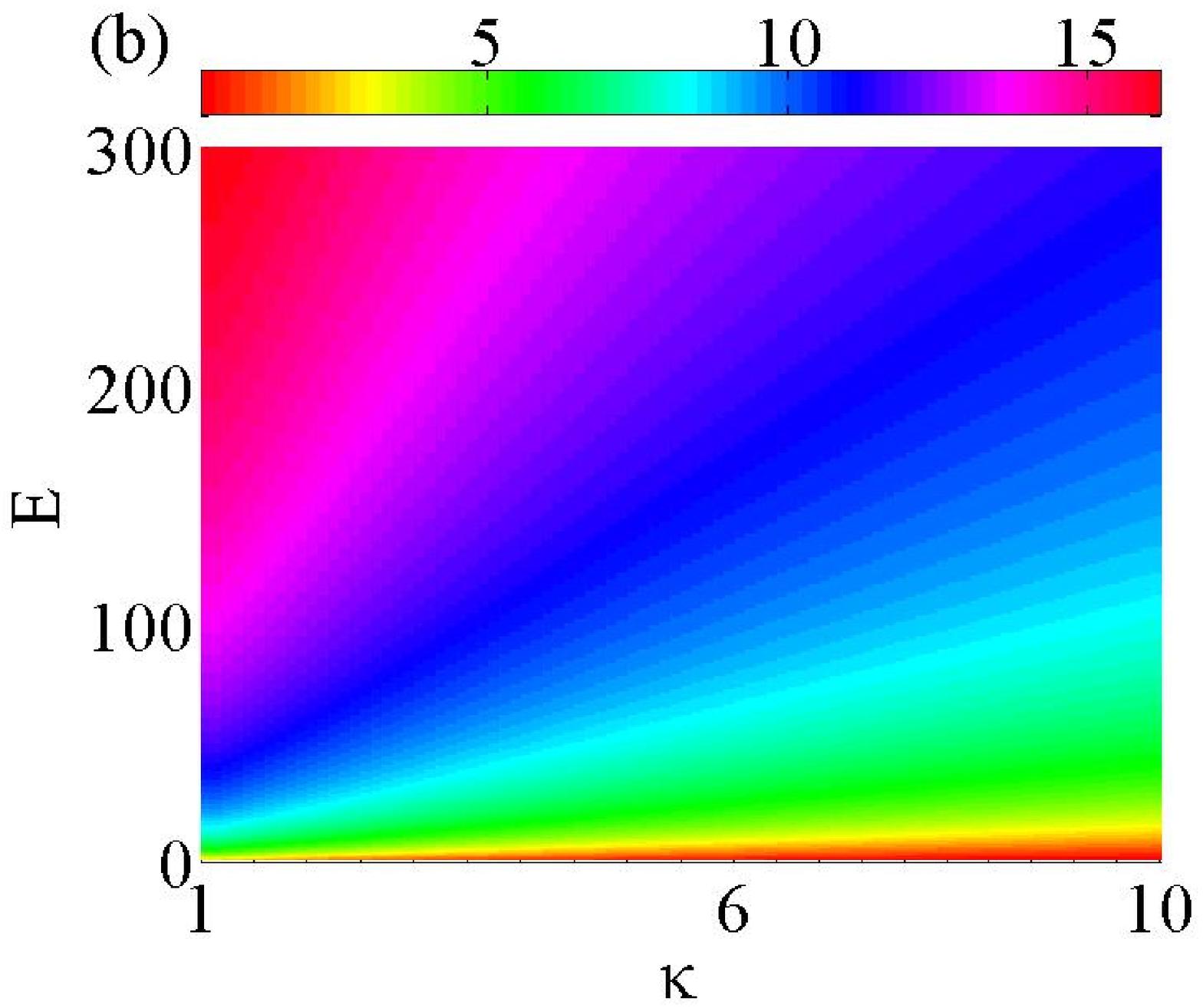}
\caption{(Color online) The quantum coherence as a function of the
driving strength $E$ and the cavity decay rate $\kappa$ (both $E$
and $\kappa$ are scaled by $\omega_{m}=1$). (a) and (b) correspond to
the mechanical coherence $C_{\textrm{mec}}$ and the optical
coherence $C_{\textrm{opt}}$, respectively. Other parameter are $\gamma_{m}=10^{-2}\,\omega_{m}$, $G_{0}=10^{-3}\,\omega_{m}$, $\Delta_{0}= \omega_{m}$, and $n_{th}=10$.}
\label{figC}
\end{figure}

\begin{figure}[!tb]
\centering
\includegraphics[width=2.5in]{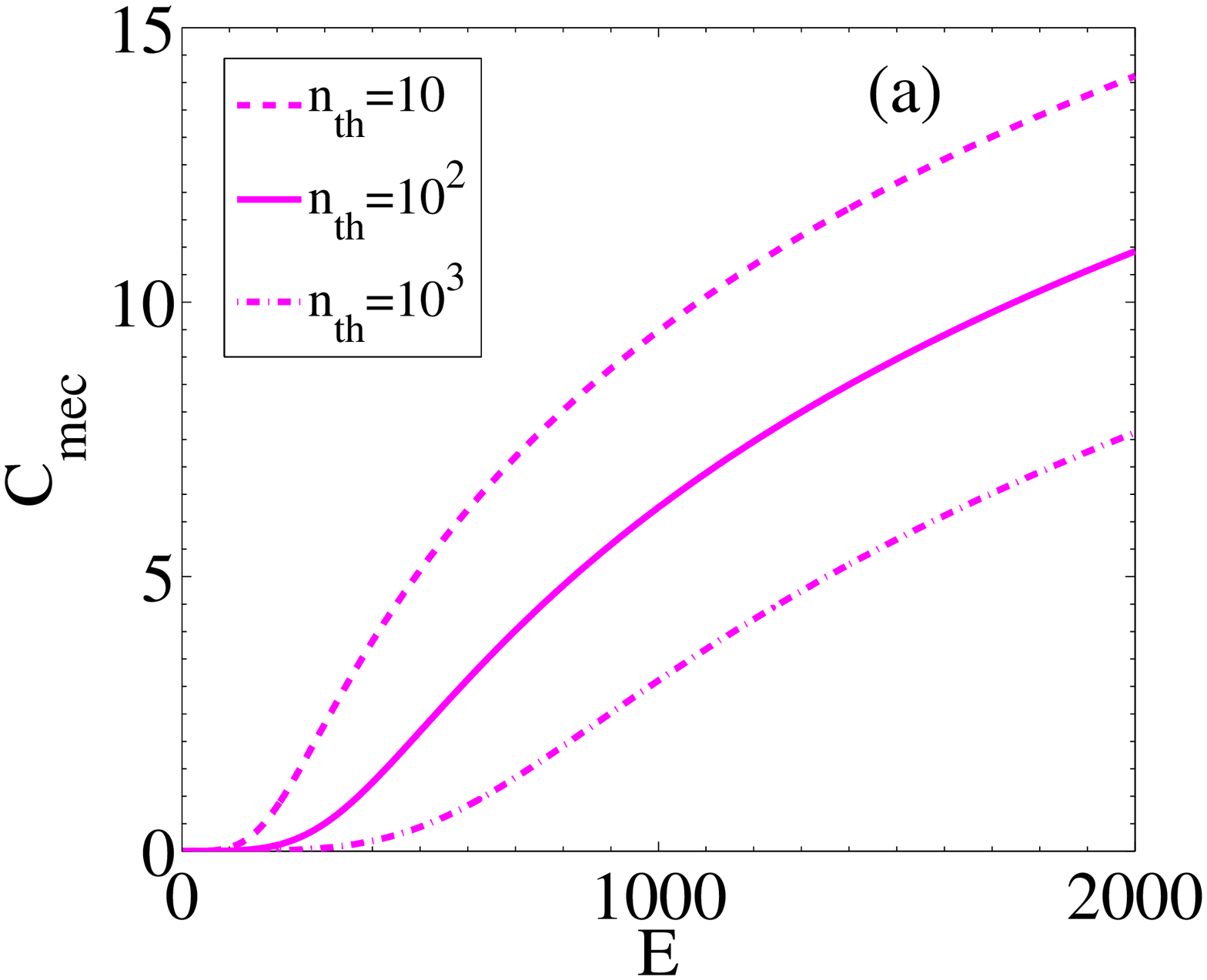}
\includegraphics[width=2.5in]{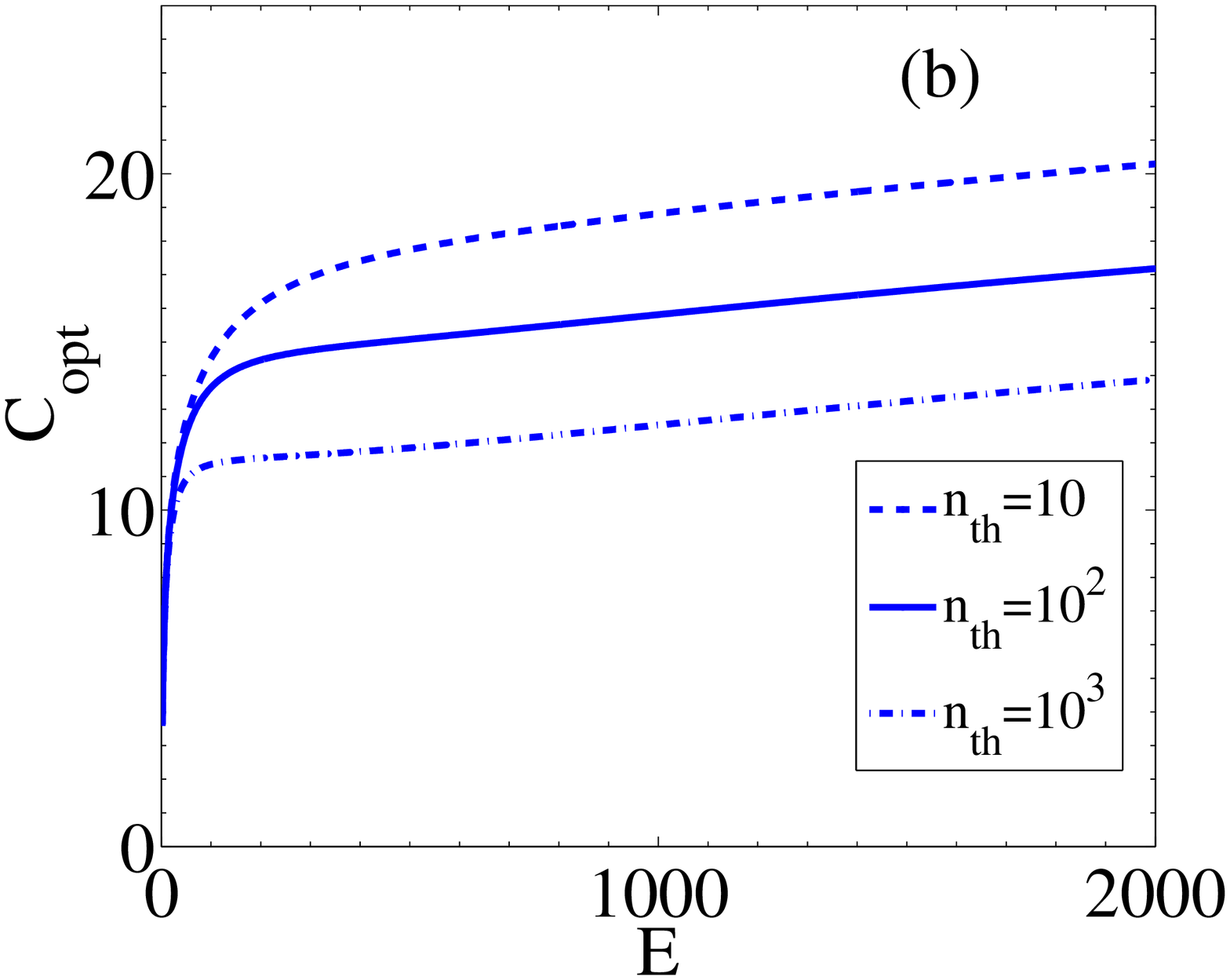}
\includegraphics[width=2.5in]{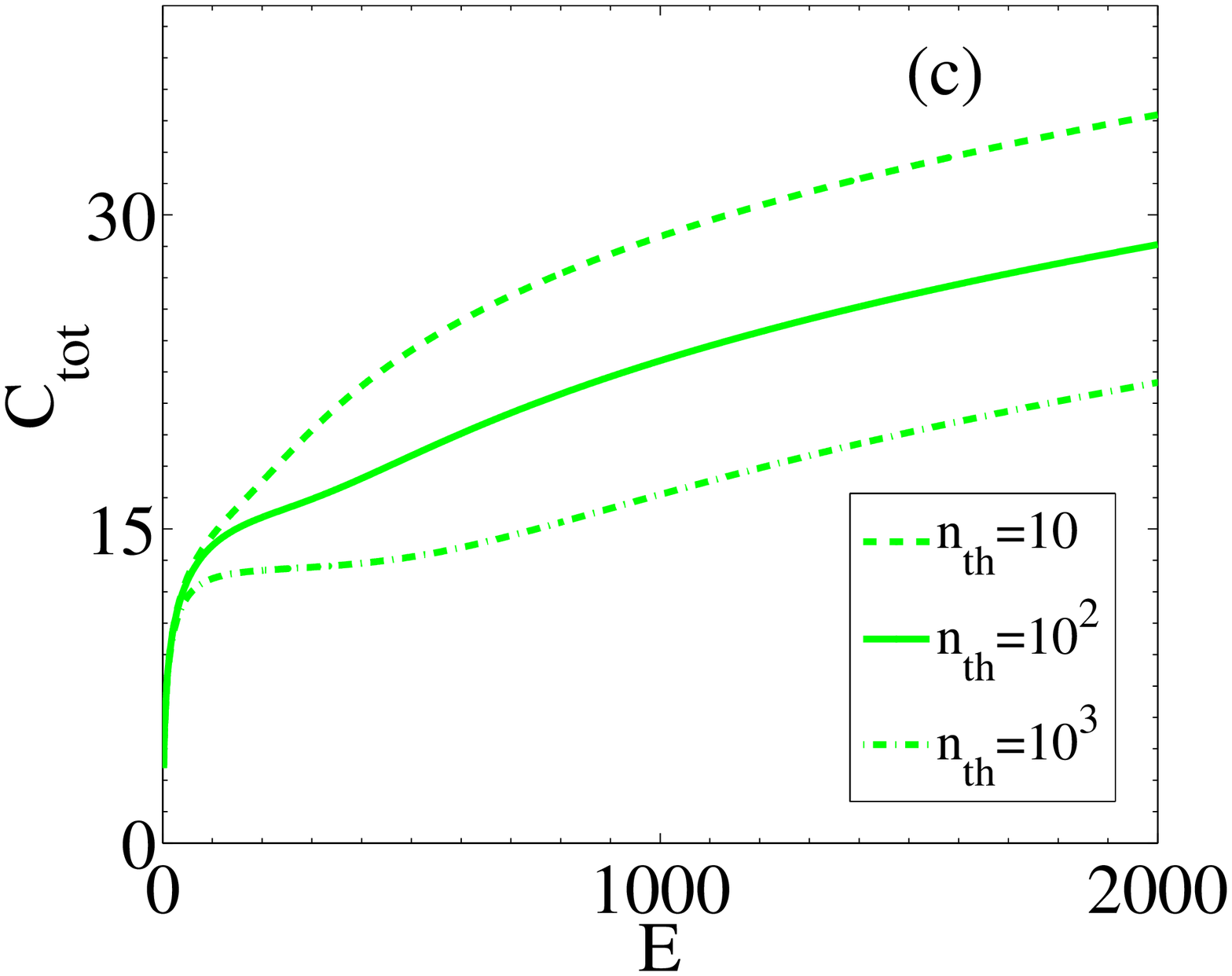}
\caption{(Color online) The quantum coherence as a function of the
driving strength $E$ (scaling with $\omega_{m}=1$) for different values of $n_{th}$. (a), (b), and (c) correspond to the mechanical coherence $C_{\textrm{mec}}$, the optical coherence $C_{\textrm{opt}}$, and the optomechanical coherence $C_{\textrm{tot}}$, respectively. Other parameter are $\gamma_{m}=10^{-2}\,\omega_{m}$, $G_{0}=10^{-4}\, \omega_{m}$, $\kappa=0.1\, \omega_{m}$, and
$\Delta_{0}= \omega_{m}$.} \label{figA}
\end{figure}

\subsection{Effect of optomechanical coupling strength}

To show the effect of the optomechanical coupling strength on the
quantum coherence, we plot in Fig.~\ref{figB} the variations of the
quantum coherence $C_{\textrm{mec}}$, $C_{\textrm{opt}}$, and
$C_{\textrm{tot}}$ in terms of the driving strength $E$ with different $G_{0}$. It is apparent that without the external optical driving, all the three kinds of coherence are null. With the increase of the strength of the driving field, the quantum coherence of the mechanical mode, the optical field, and the total optomechanical system is gradually established, respectively. And Fig.~1 clearly shows that $C_{\textrm{mec}}$, $C_{\textrm{opt}}$, and $C_{\textrm{tot}}$ increase monotonously with $E$.

Comparing Fig.~\ref{figB}(a) with Fig.~\ref{figB}(b), we find that
the relationship $C_{\textrm{opt}} \geq C_{\textrm{mec}}$ is always
valid under the same parameters. This result can be understood as following. The  optical field is subject to the zero-point fluctuation from the vacuum environment, while the mechanical subsystem is subject to the thermal noise from its bath. From this point of view, it is expected that the optical field, which is in a much less incoherent environment than that of the mechanical resonator, should have a larger degree of coherence.

It is shown in Fig.~\ref{figB}(c) that the total coherence $C_{\textrm{tot}}$ increases with the optomechanical coupling $G_{0}$. This is a reasonable result: the coherence between two subsystems of the optomechanical system can be enhanced by their strengthened coupling strength.
To our surprise, Fig.~\ref{figB}(b) displays that $C_{\textrm{opt}}$ decreases with
the increase of $G_{0}$, while $C_{\textrm{mec}}$ in
Fig.~\ref{figB}(a) goes up at the same time. Let us discuss this
counter-intuitive result more carefully. For the parameters in
Fig.~\ref{figB}, we find that the optomechanical system operates in
the red-detuned regime $\Delta_{0} \approx \omega_{m}$ with the
resolved side-band condition $\omega_{m} \gg \kappa$. In this case,
the optomechanical interaction is of the form of beam splitter in the rotating wave approximation \cite{Aspelmeyer2014a}. As a result,
the coherence should transfer from the optical mode into the
mechanical mode because $C_{\textrm{opt}} \geq C_{\textrm{mec}}$.
It is consistent with the expectation that the optical coherence
$C_{\textrm{opt}}$ should decrease with the increase of the
optomechanical coupling $G_{0}$.

We also consider the asymptotical behavior of the quantum coherence in terms of the driving field $E$. In the regime $\omega_{m} \gg G, \kappa$, the nonzero elements of the covariance matrix for the mechanical mode are given as
\cite{Vitali2007b} $\mathbf{V}_{\textrm{mec},11}=\mathbf{V}_{\textrm{mec},22} =n_{th}+\frac{1}{2}-\frac{2G^2\kappa(-n_{th}+1/2)}{(\gamma_{m}+2 \kappa )(2 \gamma_{m}\kappa+G^2)} \simeq 2 n_{th}$ within the rotation wave approximation. In the case, the quantum coherence of the mechanical resonator can be approximated as $C_{\textrm{mec}} \propto \log (G_{0}^2 E^4)$ with $\bar{n} \rightarrow 2 G_{0}^2 E^4$ in the large $E$ limit and $F(x) \rightarrow \log(x)$ in the limit $x \rightarrow \infty$ (the definitions of $\bar{n}$ and  $F(x)$ are given in the Appendix). Similarly, the coherence of the optical mode is obtained as $C_{\textrm{opt}} \propto \log (\kappa^2 E^2)$ in the same limits as that of $C_{\textrm{mec}}$. Note that in most of the realistically optomechanical systems, the single-photon coupling strength $G_{0}$ is much smaller than the cavity decay rate $\kappa$. Moreover, the driving strength $E$ must be a finite value which is restricted by the limited laser power and required by the stability of the system. Thus, for the typical parameters in the optomechanical experiments, the optical coherence $C_{\textrm{opt}}$ should be larger than the mechanical coherence $C_{\textrm{mec}}$.

\subsection{Effect of the dissipation channels}

We also explore the effect of the dissipation channels on the
optomechanical coherence. For the optomechanical system, the main
dissipation channels are the thermal damping of the mechanical
mode and the cavity decay. We find all the three kinds of quantum coherence($C_{\textrm{mec}}$, $C_{\textrm{opt}}$, and $C_{\textrm{tot}}$) are very robust to the mechanical damping rate: with the increase of the damping rate from $\gamma_{m}=10^{-3}\,\omega_{m}$ to $\gamma_{m}=10^{-2}\,\omega_{m}$, nevertheless all the values of the three kinds of coherence are suppressed, we do not observe their noticeable changes.

Fig.~\ref{figC} exhibits the role of the cavity decay on the quantum
coherence. Different from the case of $\gamma_{m}$, the cavity decay rate
$\kappa$ spoils all the three kinds of coherence obviously. Moreover,
$C_{\textrm{mec}}$ is more sensitive in terms of $\kappa$ than
$C_{\textrm{opt}}$: with the increase of the cavity decay rate from $\kappa= \omega_{m}$ to $\kappa=10\,\omega_{m}$, the value of $C_{\textrm{opt}}$ only drops half, while that of $C_{\textrm{mec}}$ decreases considerably and even
approaches zero. This result implies that the mechanical coherence
can be controlled by the optical field. The behavior of the
optomechanical coherence $C_{\textrm{tot}}$ is rather similar to
$C_{\textrm{opt}}$, so its figure is omitted here.

Fig.~\ref{figA} shows the role of the mean thermal phonon number
$n_{\textrm{th}}$ on the coherence. The behaviors of this figure are
consistent with our expectation: with the increase of
$n_{\textrm{th}}$, all the three kinds of coherence drop down. Physically, the
thermal excitation should be an incoherent operation in general.
It's interesting here that with the optomechanical coupling, the
incoherent channel of the mechanical resonator can affect the
optical coherence through the optomechanical interaction. This
result indicts that one could manipulate the coherence of a subsystem
by adjusting its remote correlated counterpart.

\begin{figure}[!tb]
\centering
\includegraphics[width=2.5in]{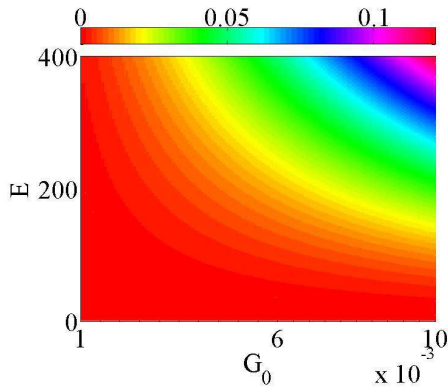}
\caption{(Color online) The quantum coherence difference $\Delta C$
as a function of the driving strength $E$ and the optomechanical-coupling strength $G_{0}$ (both $E$ and $G_{0}$ are scaled by $\omega_{m}=1$). Other parameters are $\gamma_{m}=10^{-2} \,\omega_{m}$, $\Delta_{0}=\omega_{m}$, and $n_{th}=10$.} \label{figD}
\end{figure}

\subsection{Relationship with quantum mutual entropy}

Defining $\Delta C$ as the coherence difference between the total
optomechanical system and the sum of the optical and mechanical
subsystems
\begin{equation}
\Delta C= C_{\textrm{tot}}-(C_{\textrm{mec}}+C_{\textrm{opt}} ),
\label{cohdiffdef}
\end{equation}
applying Eqs.~(\ref{incohstatA2}) and (\ref{incohstatA4}), after some straightforward calculations we find the coherence difference is just equal to the mutual information of the optomechanical system with the covariance matrix given by Eq.~(\ref{optomecCVB})
\begin{equation}
\Delta C=I(\mathbf{V}_{\textrm{tot}}).
\label{optomeCohB}
\end{equation}
Here the mutual information of a two-mode Gaussian state is defined as $I(\mathbf{V}_{\textrm{tot}})=S(\mathbf{V}_{\textrm{mec}})+S(\mathbf{V}_{\textrm{opt}}) -S(\mathbf{V}_{\textrm{tot}})$. For the optomechanical system, it can be expressed as \cite{Serafini2004}
\begin{equation}
I(\mathbf{V}_{\textrm{tot}})=F(a)+F(b)-[ F(\nu_{1})+F(\nu_{2})]
\end{equation}
with $a=\sqrt{ \mathrm{Det}(\mathbf{V_{\textrm{mec}}})}$ and $b=\sqrt{\mathrm{Det}(\mathbf{V_{\textrm{opt}}})}$.

Physically, the relationship shown in Eq.~(\ref{optomeCohB}) can be understood as following. The coherence difference $\Delta C$ defined in Eq.(\ref{cohdiffdef}) describes the inter-coherence between the optical and mechanical subsystems. And $I(\mathbf{V}_{\textrm{tot}})$ quantifies how much the optical field correlates with the mechanical mode (i.e., the area of their overlap in the Venn diagram). In this sense Eq.~(\ref{optomeCohB}) is reasonable.

In Fig.~\ref{figD}, we plot the variation of the quantum coherence
difference $\Delta C$ with respect to the intensity of the driving
laser $E$ and the optomechanical coupling strength $G_{0}$. It
displays that $\Delta C=0$ with the weak optomechanical coupling
$G_{0}< G_{0c} \simeq 3 \times 10^{-3} \omega_{m}$ for a fixed driving strength $E \simeq 300~\omega_{m}$. This implies
that although the mechanical or the optical mode has its respective coherence, there is no mutual coherence between the two subsystems. For a relatively strong optomechanical-coupling strength $G_{0}> G_{0c}$, the mutual coherence is established with $\Delta C > 0$, as shown in Fig.~\ref{figD}. This fact means that with the increase of the optomechanical coupling strength, the
correlation or coherence between the mechanical and the optical
subsystems is strengthened.

\subsection{All-optical detection of optomechanical coherence}

Even if the quantum coherence is an active subject, to our best knowledge only a few papers have explored how to detect the coherence with an experimentally available scheme \cite{Girolami2014}. In this subsection, we concentrate on this question. The difference between our paper and Ref.~\cite{Girolami2014} lies in: our optomechanical system is an infinite-dimensional system, while
that studied in Ref.~\cite{Girolami2014} is a finite one.

The state of the optomechanical system is Gaussian, which only requires
the first two moments to completely describe it. Fortunately,
the optomechanical Gaussian state can be fully reconstructed by the
high-precision all-optical detection method. With these moments, we are able to
obtain the optomechanical coherence according to Eqs.~(\ref{incohstatA4}). We mainly study how to detect the second moments of the cavity optomechanical system, as its first moments are determined by
Eqs.~(\ref{moeAB8}).

For the optical subsystem, its second moments can be obtained by
the homodyne detection of the output field, which is a standard route in quantum optics~\cite{Leonhardt05}. For the subsystem of the mechanical resonator, we exploit a similar measurement scheme discussed in Ref.~\cite{Vitali07}, which has been realized in a recent experiment~\cite{Teufel2013}. With the assumption that the mechanical resonator is perfectly reflective at both sides, we add an additional fixed mirror adjacent to the other side of the mechanical resonator to form a second Fabry-P$\acute{e}$rot cavity which also couples to the mechanical resonator. Analogous to Eq.~(\ref{moeA}c), the equation of motion for the fluctuational part $\delta \hat{c}$ of the second cavity is written as
\begin{equation}
\frac{d }{dt} \delta \hat{c} = -(\kappa_{2}+i \Delta_{2}) \delta
\hat{c}+ iG_{2}( \delta \hat{b} + \delta \hat{b}^{\dag}
)/\sqrt{2}+\sqrt{2\kappa_{2}} \delta \hat{c}_{\textrm{in}},
\label{detectA}
\end{equation}%
where $\kappa_{2}$, $\Delta_{2}$, and $\delta \hat{c}_{\textrm{in}}$
are the decay rate, the effective detuning, and the input noise of
the second cavity mode, respectively; $G_{2}$ is the related effective
optomechanical coupling strength. Assuming that $\langle \hat{a}
\rangle_{s} \gg \langle \hat{c} \rangle_{s}$ and $\Delta_{2}=\omega_{m} \gg \{\kappa_{2}, G_{2}\}$, which means the rotating wave approximation is fulfilled, Eq.~(\ref{detectA}) will reduce to (in the rotating frame with respect to $\omega_{m}$)
\begin{equation}
\frac{d }{dt} \delta \hat{c} = -\kappa_{2} \delta \hat{c}+ i G_{2}
\delta \hat{b} /\sqrt{2}+\sqrt{2\kappa_{2}} \delta
\hat{c}_{\textrm{in}} \label{detectB}
\end{equation}
under the rotating wave approximation. With the bad-cavity condition
$\kappa_{2} \gg  G_{2}$, the fluctuation of the output field of the second cavity is given as
\begin{equation}
\delta \hat{c}_{\textrm{out}} = \frac{i G_{2}}{\sqrt{2 \kappa_{2}} }
\delta \hat{b} + \delta \hat{c}_{\textrm{in}}. \label{detectC}
\end{equation}
Eq.~(\ref{detectC}) indicts that the second moments of the mechanical subsystem can be obtained by homodyne measurement of the second cavity. And we can reconstruct the correlational matrix of the optical and mechanical subsystems through measuring the moment correlations of the outputs from the two cavities.

Last but not least, the real values of the experimental parameters in two groups \cite{Teufel2016, chan2011} are given to show the feasibility of our detection scheme. For the optomechanical coupling in the microwave domain achieved in NIST group \cite{Teufel2016}, the mechanical mode has the resonance frequency $\omega_{m}/2\pi=14.98$\,MHz with an intrinsic damping rate $\gamma_{m}/2\pi=9.2$\,Hz. The microwave cavity has the resonance frequency $\omega_{0}/2\pi=8.89$\,GHz and line-width $\kappa/2\pi=1.17$\,MHz. The single-photon optomechanical coupling strength $G_{0}=145$\,Hz. In the optomechanical crystal system~\cite{chan2011}, the corresponding parameters are $\omega_{m}/2\pi=3.68$\,GHz, $\gamma_{m}/2\pi=35$\,kHz, $\omega_{0}/2\pi=195$\,THz, $\kappa/2\pi=500$\,MHz, and $G_{0}/2\pi=910$\,kHz. It's clear that the parameters used in our measurement scheme for optomechanical coherence has the same order of magnitude as that of the two actually experimental implementation, and the following conditions $\omega_{m} \gg \kappa$ and $\kappa \gg G_{0}$ are satisfied.

\section{Conclusion}

In summery, we explore the macroscopic quantum coherence of the cavity optomechanical system. Using the standard linearized approximation, the first and second moments of the fluctuations of the optomechanical system in the steady state are obtained. We show that the external optical driving laser can enhance the macroscopic quantum coherence of the cavity optomechanics, while the optical and mechanical dissipations suppress the optomechanical coherence. In addition, we also find that the difference between the total optomechanical coherence and the sum of the coherence of the optical and mechanical subsystems just equals their quantum mutual information. Moreover, an all-optical detection scheme is suggested to experimentally probe the macroscopic optomechanical coherence.

\begin{acknowledgments}
Q. Zheng thank G. Adesso for his insightful comments and discussions. This work is supported by the National Natural Science Foundation of China (Grants No. 11365006, No. 11347213, No. 11565010, No. 11422437, and No. U1530401), the 973 program (Grants No. 2014CB921403 and No. 2016YFA0301200), and the National Natural Science Foundation of Guizhou Province QKHLHZ[2015]7767.
\end{acknowledgments}

\appendix

\section{Quantum coherence as resource}

Let us first discuss the quantum coherence for a system with a finite-dimensional Hilbert space. In a pre-fixed finite basis $\{ |k\rangle \}$ ($k=0, 1,..., D$), any incoherent state can be written as
$\sigma = \sum_{k}\sigma_{k} |k\rangle \langle k|$. Here $\sigma \in \mathfrak{I}$ with $ \mathfrak{I}$ being a set of incoherent states. Ref.~\cite{Plenio2014} suggests that any proper measure of quantum coherence $C$ in a finite Hilbert space must obey the following postulates: (a) $C(\rho) \geq 0$ for any quantum state $\rho$ and $C(\rho)= 0$ iff $\rho \in \mathfrak{I}$; (b1) $C$ is monotonic under all incoherent completely positive and trace-preserving maps $\Phi$:
$C(\rho)\geq C(\Phi(\rho))$; (b2) $C$ is monotonic for average coherence under subselection based on measurement outcomes; (c) $C$ is nonincreasing under the classical mixing of quantum states.

For a finite-dimensional density matrix $\rho=\sum_{i, j}\rho_{ij} |i\rangle \langle j|$, based on the relative entropy $S(\rho||\sigma)=\mathrm{Tr}[\rho \mathrm{log}(\rho)]- \mathrm{Tr}[\rho \mathrm{log}(\sigma)]$, the measure of coherence is given as \cite{Plenio2014}
$C_{\textrm{ent}}=\mathrm{min}_{\sigma \in \mathfrak{I}} S(\rho||\sigma)
=S(\rho_{\textrm{diag}})- S(\rho)$, where $S(\rho)=-\mathrm{Tr}(\rho \mathrm{log} \rho)$ is von-Neumman entropy, $\rho_{\textrm{diag}}=\sum_{i}\rho_{ii} |i\rangle \langle i|$ as the corresponding diagonal density matrix. It has been proved that this coherence quantifiers fulfill the above mentioned conditions (a), (b1), (b2) and (c).

Some relevant physical situations, such as quantum state of light, need the coherence theory of
infinite-dimensional system. Up to now, this tricky question at
least has been investigated by two groups in Refs. \cite{yzhanghf2016, JWXU16}.
In this paper, we will follow the framework of Ref.~\cite{JWXU16} to study the coherence of the cavity optomechanical system. The benefit of Ref.~\cite{JWXU16} is that the coherence for Gaussian state is expressed in the closed-form in terms of the covariance matrices and displacement vectors. This makes the related calculation relatively easy in practice as there is no need to cut off the Hilbert space to the finite dimension. And more importantly, it can be detected experimentally with the state of art set-up (more details are given in Sec. III D).

Now, we investigate the coherence of a single-mode Gaussian state.
For a given bosonic mode $\hat{b}$ with the commutation relation
$[\hat{b}, \hat{b}^{\dag}]=1$, we can define the quadrature operators
\begin{equation}
\hat{q} = \hat{b}+ \hat{b}^{\dag},~~\hat{p} = i(\hat{b}^{\dag}-\hat{b})
\end{equation}
supported in the Fock space $ \{ | n \rangle \}$.
Denoting the collective operator $\hat{\vec{x}}=(\hat{q}, \hat{p})$,
for a Gaussian state $\rho$, it can be fully described by its first moment $\vec{d}=(d_{1}, d_{2})= \mathrm{Tr}(\rho \hat{\vec{x}}) $ and second moment (the so-called covariance matrix) $\mathbf{V}=\left(
\begin{array}{cc}
V_{11} & V_{12} \\
V_{21} & V_{22} \\
\end{array}%
\right)$ with the elements defined as $V_{ij}=\langle \hat{x}_{i}\hat{x}_{j}+\hat{x}_{j}\hat{x}_{i} \rangle/2-\langle \hat{x}_{i}\rangle \langle \hat{x}_{j}\rangle$.

In Appendix A of Ref.~\cite{JWXU16}, it has been proved that a
one-mode Gaussian state is diagonal if and only if it is a
thermal state in the Fock space. Based on this result, the coherence of any given one-mode Gaussian state $\rho(\mathbf{V}, \vec{d})$ is defined as \cite{JWXU16}
\begin{equation}
C[\rho(\mathbf{V}, \vec{d})]= \mathrm{inf}_{\delta }~{ S( \rho(\mathbf{V}, \vec{d})|| \delta)}
=-F(\nu)+F(2\bar{n}+1 ).
\label{incohstatA2}
\end{equation}
Here $\delta$ denotes an incoherent Gaussian state,
$\nu = \sqrt{ \mathrm{Det}(\mathbf{V})}$,
$\bar{n} = (V_{11}+V_{22}+d_{1}^2+d_{2}^2-2)/4$,
and
$F(x)=\frac{x+1}{2}\mathrm{log}( \frac{x+1}{2})- \frac{x-1}{2}\mathrm{log}( \frac{x-1}{2})$.
Note that this measure is only valid for Gaussian state under the operation of the (incoherent) Gaussian channel. And the conditions of coherence (b1) and (b2) for finite-dimensional case are replaced by the following one \cite{JWXU16}:
$C(\rho)\geq C(\Phi_{\textrm{IGC}}(\rho))$ for any Gaussian state and incoherent Gaussian channel (IGC).

It is straightforward to generalize the above result to
multi-mode Gaussian state. Here we are only interested in the
two-mode Gaussian state, which is most relevant to the cavity optomechanical system to be
discussed in Sec. III. The coherence of an arbitrary two-mode
Gaussian state is given as \cite{JWXU16}
\begin{equation}
C[\rho(\mathbf{V}, \vec{d})] = \sum_{i=1}^{2} [-F(\nu_{i})+
F(2\bar{n}_{i} +1 ) ]
\label{incohstatA4}
\end{equation}
with $\nu_{i}$ ($i=1, 2$) is the symplectic eigenvalue of $\mathbf{V}$ and
$\bar{n}_{i}=(V_{11}^{(i)}+V_{22}^{(i)}+(d_{1}^{(i)})^2+(d_{2}^{(i)})^2-2)/4$ is
determined by the $i$th-mode covariance matrix
$\mathbf{V}^{(i)}$ and displacement vector $\vec{d}^{(i)}$.




\begin{thebibliography}{99}

\bibitem{Dirac58}
P. A. M. Dirac, \textit{The Principles of Quantum Mechanics}
(Oxford University Press, Oxford, 1958).



\bibitem{JJMA2016}
J. J. Ma, B. Yadin, D. Girolami, V. Vedral, and M. Gu, Phys. Rev. Lett. \textbf{116}, 160407 (2016).

\bibitem{yao2015}
Y. Yao, X. Xiao, L. Ge, and C. P. Sun, Phys. Rev. A \textbf{92}, 022112 (2015).

\bibitem{Streltsov2015}
A. Streltsov, U. Singh, H. S. Dhar, M. N. Bera, and G.
Adesso, Phys. Rev. Lett. \textbf{115}, 020403 (2015).


\bibitem{Nielsen00}
M. Nielsen and I. Chuang, \textit{Quantum Computation
and Quantum Information} (Cambridge University Press,
Cambridge, England, 2000).



\bibitem{Maccone2014}
R. Demkowicz-Dobrzanski and L. Maccone, Phys. Rev.
Lett. \textbf{113}, 250801 (2014).


\bibitem{Correa2014}
L. A. Correa, J. P. Palao, D. Alonso, and G. Adesso, Sci.
Rep. \textbf{4}, 3949 (2014); J. {\AA}berg, Phys. Rev. Lett. \textbf{113}, 150402 (2014);
M. Lostaglio, D. Jennings, and T. Rudolph, Nat. Commun.
\textbf{6}, 6383 (2015).


\bibitem{SWLISUN0215}
S. W. Li, C. Y. Cai, C. P. Sun, Ann. Phys. \textbf{360}, 19 (2015).



\bibitem{Collini2010}
E. Collini, C. Y. Wong, K. E. Wilk, P. M. G. Curmi, P.
Brumer, and G. D. Scholes, Nature (London) \textbf{463}, 644
(2010).



\bibitem{Spekkens2014}
I. Marvian and R. W. Spekkens, Nat. Commun. \textbf{5}, 3821 (2014).



\bibitem{Glauber63}
R. J. Glauber, Phys. Rev. \textbf{131}, 2766 (1963).


\bibitem{Sudarshan63}
E. C. G. Sudarshan, Phys. Rev. Lett. \textbf{10}, 277 (1963).


\bibitem{Plenio2014}
T. Baumgratz, M. Cramer, and M. B. Plenio, Phys. Rev. Lett. \textbf{113}, 140401 (2014).

\bibitem{Mintert2014}
F. Levi and F. Mintert, New J. Phys. \textbf{16}, 033007 (2014).


\bibitem{Marvian2013}
I. Marvian and R. W. Spekkens, New J. Phys. \textbf{15}, 033001 (2013).



\bibitem{linori2012}
C.-M. Li, N. Lambert, Y.-N. Chen, G.-Y. Chen, and F. Nori,
Sci. Rep. \textbf{2}, 885 (2012).


\bibitem{guolilam2015}
Z. Xi, Y. Li, and H. Fan, Sci. Rep. \textbf{5}, 10922 (2015);
S. Du, Z. Bai, and Y. Guo, Phys. Rev. A \textbf{91}, 052120 (2015);
X. Yuan, H. Zhou, Z. Cao, and X. Ma, {\it ibid}. \textbf{92},
022124 (2015); Y. Peng, Y. Jiang, and H. Fan, {\it ibid}. \textbf{93}, 032326 (2016).



\bibitem{Winter2016}
A. Winter and D. Yang, Phys. Rev. Lett. \textbf{116}, 120404 (2016).



\bibitem{Adesso2016b}
E. Chitambar, A. Streltsov, S. Rana, M. N. Bera, G. Adesso,
and M. Lewenstein, Phys. Rev. Lett. \textbf{116}, 070402 (2016);
T. R. Bromley, M. Cianciaruso, and G. Adesso, Phys. Rev.
Lett. \textbf{114}, 210401 (2015).


\bibitem{vedral2016}
B. Yadin and V. Vedral, Phys. Rev. A \textbf{93}, 022122 (2016).



\bibitem{Parthasarathy2016}
C. Radhakrishnan, M. Parthasarathy, S. Jambulingam, and T. Byrnes, Phys. Rev.
Lett. \textbf{116}, 150504 (2016); E. Bagan, J. A. Bergou, S. S. Cottrell, and M. Hillery, Phys. Rev. Lett. \textbf{116}, 160406 (2016).



\bibitem{luo2012}
S. Luo, S. Fu, and C. H. Oh, Phys. Rev. A \textbf{85}, 032117 (2012).

\bibitem{Rastegin2016}
A. E. Rastegin, Phys. Rev. A \textbf{93}, 032136 (2016).




\bibitem{Anderson1995}
M. H. Anderson, J. R. Ensher, M. R. Matthews, C. E. Wieman,
and E. A. Cornell, Science \textbf{269}, 198 (1995); K. B. Davis, M.
O. Mewes, M. R. Andrews, N. J. van Druten, D. S. Durfee, D.
M. Kurn, and W. Ketterle, Phys. Rev. Lett. \textbf{75}, 3969 (1995).


\bibitem{younori2005}
J. Q. You, F. Nori, Phys. Today \textbf{58}, 42 (2005).


\bibitem{Cleland2010}
A. D. OConnell, M. Hofheinz, M. Ansmann, R. C. Bialczak, M. Lenander, E. Lucero, M. Neeley, D. Sank, H. Wang, M. Weides, J. Wenner, J. M. Martinis, and A. N. Cleland, Nature \textbf{464}, 697 (2010).



\bibitem{Aspelmeyer2012a}
M. Aspelmeyer, P. Meystre, and K. C. Schwab, Phys. Today
\textbf{65}, 29 (2012).


\bibitem{Aspelmeyer2014a}
M. Aspelmeyer, T. J. Kippenberg, and F. Marquardt, Rev. Mod.
Phys. \textbf{86}, 1391 (2014).



\bibitem{Rae2007}
I. Wilson-Rae, N. Nooshi, W. Zwerger, and T. J. Kippenberg,
Phys. Rev. Lett. \textbf{99}, 093901 (2007); F. Marquardt, J. P. Chen,
A. A. Clerk, and S. M. Girvin, {\it ibid}. \textbf{99}, 093902 (2007).



\bibitem{LTYDWANG2013}
L. Tian, Phys. Rev. Lett. \textbf{110}, 233602 (2013); Y. D. Wang and
A. A. Clerk, \textit{ibid}. \textbf{110}, 253601 (2013).


\bibitem{Barzanjeh2012a}
Sh. Barzanjeh, M. Abdi, G. J. Milburn, P. Tombesi, and D. Vitali,
Phys. Rev. Lett. \textbf{109}, 130503 (2012).



\bibitem{Kronwald2013}
A. Kronwald, F. Marquardt, A. A. Clerk, Phys. Rev. A \textbf{88},
063833 (2013).



\bibitem{Wollman2015}
E. E. Wollman, C. U. Lei, A. J. Weinstein, J. Suh, A. Kronwald, F. Marquardt,
A. A. Clerk, K. C. Schwab, Science, \textbf{349}, 952 (2015).



\bibitem{Agarwal2015}
K. Qu and G. S. Agarwal, Phys. Rev. A \textbf{91}, 063815 (2015).


\bibitem{Painter2012}
A. H. Safavi-Naeini, J. Chan, J. T. Hill, Thiago P. Mayer Alegre,
A. Krause, and O. Painter, Phys. Rev. Lett. \textbf{108}, 033602 (2012).


\bibitem{purdy2013}
T. P. Purdy, R. W. Peterson and C. A. Regal, Science \textbf{339}, 801 (2013);
J. D. Teufel, F. Lecocq and R. W. Simmonds, Phys. Rev. Lett. \textbf{116}, 013602 (2016).




\bibitem{jqliao2016}
J. Q. Liao and L. Tian, Phys. Rev. Lett. \textbf{116}, 163602 (2016).


\bibitem{shbar15}
Sh. Barzanjeh, S. Guha, C. Weedbrook, D. Vitali, J. H.
Shapiro, and S. Pirandola, Phys. Rev. Lett. \textbf{114}, 080503 (2015).


\bibitem{Zhang2014a}
K. Zhang, F. Bariani, and P. Meystre, Phys. Rev. Lett. \textbf{112}, 150602 (2014).



\bibitem{walls2008}
D. F. Walls and G. J. Milburn, \textit{Quantum Optics} (Springer,
Berlin, 2008).



\bibitem{pzoller00}
C. W. Gardiner and P. Zoller, \textit{Quantum Noise} (Springer, Berlin,
2000).


\bibitem{Clerk2011}
A. A. Clerk, M. H. Devoret, S. M. Girvin, F. Marquardt, and
R. J. Schoelkopf, Rev. Mod. Phys. \textbf{82}, 1155 (2010).





\bibitem{Benguria1981}
R. Benguria, and M. Kac, Phys. Rev. Lett, \textbf{46}, 1 (1981).



\bibitem{DengXTX2016}
Z. J. Deng, X. B. Yan, Y. D. Wang, and C. W. Wu, Phys. Rev. A \textbf{93}, 033842 (2016);
W. Xiong, D. Jin, Y. Qiu, C. H. Lam, and J. Q. You, {\it ibid}. \textbf{93}, 023844 (2016);
H. T. Tan, F. Bariani, G. X. Li, and P. Meystre, \textit{ibid}. \textbf{88}, 023817 (2013).



\bibitem{yli2012a}
W. J. Nie, Y. H. Lan, Y. Li, and S. Y. Zhu, Phys. Rev. A \textbf{86}, 063809 (2012).



\bibitem{xuli2015}
X. W. Xu and Y. Li, Phys. Rev. A \textbf{91}, 053854 (2015).


\bibitem{Meystre2008}
M. Bhattacharya, P. L. Giscard, and P. Meystre, Phys. Rev. A \textbf{77}, 013827 (2008).



\bibitem{DeJesus1987}
E. X. DeJesus and C. Kaufman, Phys. Rev. A \textbf{35}, 5288(1987).



\bibitem{Plenio2008}
M. J. Hartmann and M. B. Plenio, Phys. Rev. Lett. \textbf{101}, 200503 (2008).



\bibitem{Paternostro2007}
M. Paternostro, D. Vitali, S. Gigan, M. Kim, C. Brukner, J. Eisert, and M. Aspelmeyer, Phys. Rev. Lett. \textbf{99}, 250401 (2007).



\bibitem{Vitali2007b}
D. Vitali, P. Tombesi, M. J. Woolley, A. C. Doherty, and G. J. Milburn,
Phys. Rev. A \textbf{76}, 042336 (2007).



\bibitem{Serafini2004}
A. Serafini, F. Illuminati and S. Siena, J. Phys. B \textbf{37}, L21 (2004).



\bibitem{Girolami2014}
D. Girolami, Phys. Rev. Lett. \textbf{113}, 170401 (2014).





\bibitem{Leonhardt05}
U. Leonhardt, \textit{Measuring the Quantum State of Light} (Cambridge Univ.
Press, 2005).



\bibitem{Vitali07}
D. Vitali, S. Gigan, A. Ferreira, H. R. B$\ddot{o}$hm, P. Tombesi,
A. Guerreiro, V. Vedral, A. Zeilinger, and M. Aspelmeyer, Phys.
Rev. Lett. \textbf{98}, 030405 (2007).



\bibitem{Teufel2013}
T. A. Palomaki, J. D. Teufel, R. W. Simmonds, and K. W. Lehnert, Science \textbf{342}, 710 (2013).




\bibitem{Teufel2016}
F. Lecocq, J. B. Clark, R.W. Simmonds, J. Aumentado, and J. D. Teufel,
Phys. Rev. Lett. \textbf{116}, 043601 (2016).


\bibitem{chan2011}
J. Chan, T. P. Mayer Alegre, A. H. Safavi-Naeini, J. T. Hill, A.
Krause, S. Groblacher, M. Aspelmeyer, and O. Painter, Nature
\textbf{478}, 89 (2011).




\bibitem{yzhanghf2016}
Y. Zhang, L. Shao, Y. Li, and H. Fan, Phys. Rev. A \textbf{93}, 012334 (2016).


\bibitem{JWXU16}
J. Xu, Phys. Rev. A \textbf{93}, 032111 (2016).


\bibitem{Weedbrook2012}
C. Weedbrook, S. Pirandola, R. Garcia-Patron, N. J. Cerf,
T. C. Ralph, J. H. Shapiro, and S. Lloyd, Rev. Mod. Phys. \textbf{84}, 621
(2012).



\bibitem{Ferraro2005}
A. Ferraro, S. Olivares, and M. G. A. Paris, \textit{Gaussian States
in Quantum Information}, Napoli Series on Physics and Astrophysics
(Bibliopolis, Napoli, 2005).




\end{thebibliography}
\end{document}